# Inversion of exciton level splitting in quantum dots

R. J. Young[†][1,2], R. M. Stevenson[1], A. J. Shields[1], P. Atkinson[2], K. Cooper[2], D. A. Ritchie[2], K. M. Groom[3], A. I. Tartakovskii[3], and M. S. Skolnick[3]

[1]Toshiba Research Europe Limited, 260 Cambridge Science Park. Cambridge CB4 0WE, UK
[2]Cavendish Laboratory, University of Cambridge, Madingley Road, Cambridge CB3 0HE, UK
[3]Department of Physics and Astronomy, University of Sheffield, Hounsfield Road, Sheffield S3 7RH, UK

**Abstract.**

The demonstration of degeneracy of the exciton spin states is an important step towards the production of entangled photons pairs from the biexciton cascade. We measure the fine structure of exciton and biexciton states for a large number of single InAs quantum dots in a GaAs matrix; the energetic splitting of the horizontally and vertically polarised components of the exciton doublet is shown to decrease as the exciton confinement decreases, crucially passing through zero and changing sign. Thermal annealing is shown to reduce the exciton confinement, thereby increasing the number of dots with splitting close to zero.

[†] Corresponding author, ry211@cam.ac.uk

Single InAs quantum dots have attracted escalating interest recently due to their suitability as the basis for photon emission technology for applications in quantum information. Much progress has been made, particularly in the field of single photon emission devices[1-3], which have been demonstrated with electrical injection[4], high efficiency[5], and strong Purcell enhancement of the spontaneous emission rate[6-9]. It has been proposed that the decay of the biexciton state to the ground state could be used to create pairs of photons[10] entangled by their polarisation[11], due to the superposition of the two alternate decay paths, distinguished by the order in which the spin up and spin down excitons recombine. However, polarised cross-correlation measurements on the exciton and biexciton photons have confirmed that a splitting of the intermediate exciton state results in spectral 'which path' information, and the emission of only polarisation correlated photon pairs[12-14]. In this letter we present the first report of the reduction of this exciton level splitting through the zero point, a crucial step towards the generation of entangled photon pairs.

The splitting of the exciton level has been measured in many dot systems including; InAs/GaAs[12], CdSe/ZnSe[15], InP/GaInP[16], and interface fluctuations in thin GaAs/AlGaAs quantum wells[17]. In such structures, the electron-hole exchange interaction is anisotropic in the plane, due to corresponding in-plane asymmetries in the carrier wavefunctions. Elongation of the base of the quantum dot, and effects on the confinement potentials due to strain are both thought to contribute to the anisotropy. The exchange interaction leads to the hybridisation and lifting of the degeneracy of the two optically active electron – heavy hole exciton states of total angular momentum $m=\pm 1$. The splitting of these radiative exciton states is thus dependent on both the symmetry, and overlap of the constituent electron and hole wavefunctions.

The radiative decay of the biexciton state is illustrated in the level diagram inset to Fig. 1. Since the biexciton state is spin neutral, two radiative decay paths to the ground state are possible, characterised by the photon polarisation and intermediate exciton state, and indicated by arrows. The exciton level splitting, labelled S, can therefore be determined experimentally from polarisation dependent spectra of the biexciton or exciton emission lines. Control of the exciton level splitting has been demonstrated by two groups[18,19], both of which measured the splitting in the time domain, on ensembles of quantum dots. These initial results have been very encouraging, with observation of a clear reduction of the splitting, although without achieving a reduction to zero.

In contrast to previous work where average splittings of dot ensembles are measured, we measure the polarised emission lines of individual dots with high precision (~3μeV), allowing the magnitude and sign of the splitting to be probed. Although there is a wide variation in splitting within a single sample we find that there is a good correlation with the emission energy of the dot. We demonstrate that this can be controlled by the natural variation across a wafer as grown[20], or by annealing to blue-shift the emission[18]. We attribute the inversion of the exciton splitting to delocalisation of its wavefunction, which can be probed by measuring the diamagnetic shift.

The samples used for all measurements presented contained a single low density (<1μm$^{-2}$) layer of self assembled InAs quantum dots grown by MBE in a GaAs matrix[3]. One sample was left un-annealed and six thermally annealed for; 5 and 10 minutes at 650°C, 5 and 15 minutes at 700°C and 5 and 10 minutes at 750°C respectively. Single quantum dots were isolated by 2μm diameter apertures in gold films, deposited on the surface of the samples after annealing. The samples were cooled to ~5K using a continuous flow helium cryostat. A 532nm CW laser was focused through an infinity-corrected microscope objective lens to a spot size of ~1μm, and used to excite the sample above the GaAs band gap. Photoluminescence (PL) from the sample was collimated by the same lens, dispersed by a grating spectrometer and detected using a charge-couple device (CCD) camera. Neutral exciton (X) and biexciton (XX) states from single quantum dots were identified and distinguished from charged exciton states by measuring the power dependence of the spectra, and the fine structure splitting where possible[21].

Fig. 1a shows typical polarised PL spectra taken from a single quantum dot on the un-annealed sample. Two pairs of lines are visible in the spectrum, one corresponding to emission from the exciton state and the other from the biexciton state. Each doublet consists of a pair of orthogonally linearly polarised lines, equally separated, by 42 ± 3 μeV in this example, due to the exchange induced lifting of the degeneracy of the exciton. The lower energy X line, and the higher energy XX line, are predominantly polarised along the [1-10] direction of the crystal (vertically in the lab frame), in agreement with other studies[12], indicating an elongation of the exciton envelope along the [1-10] direction. The measured line widths are broader than our resolution limit of ~20μeV and are thought to be broadened by a fluctuating charge distribution close to the dot[22].

The spectra in Fig. 1b show polarised PL from a typical single quantum dot from the sample annealed for five minutes at 750°C. In these spectra both the binding energy[18] and the exciton (biexciton) fine structure splitting $E_{HX} - E_{VX}$ ($E_{VXX} - E_{HXX}$), -19 ± 3 μeV, are of different sign to those observed in Fig. 1a, where the energy of the exciton and biexciton lines are denoted by the subscripts *X* and *XX* respectively, and the vertical and horizontally polarised components are denoted by *V* and *H*.

To minimise systematic errors, and accurately measure the exciton level splitting *S*, the exciton and biexciton emission lines were fitted using Lorentzian line shapes, and the splitting energy was determined following the relationship $S = [(E_{HX} - E_{VX}) + (E_{VXX} - E_{HXX})]/2$. The energy *S* is thus positive when the horizontally polarised [110] exciton line lies to higher energy. Repeated measurements in different cool down cycles of the system showed that the splittings for most of the dots were reproducible to within 2.1μeV, as shown in the inset to Fig. 2. We can therefore expect S to be measured to an accuracy of at least ~3μeV, though estimates of the precision within a single cool down cycle are actually much better, within 0.5μeV

Fig. 2 shows the splitting of the optically active exciton state *S* versus the exciton emission energy for all the samples. A strong correlation between emission energy and splitting is clearly observed, indicated by the linear fit to the data shown as a guide to the eye. The splitting is seen to decrease from ~80μeV for the lowest energy dots, to zero as the dot energy increases to 1.4eV. To higher energy the trend continues, and the splitting becomes inverted for the highest energy quantum dots. This result demonstrates that the exciton envelope has begun to elongate along the [110] direction, orthogonal to the direction observed in previous studies. Furthermore, dots annealed for longer times and at high temperatures tend to have higher emission energies due to intermixing, and thus smaller splittings, in agreement with other annealing studies[18,19]. However, dots from the un-annealed sample cover almost the entire energy range achieved on annealed samples, and show precisely the same correlation between splitting and emission energy. We therefore conclude that this result is a general one, depends on the exciton recombination energy and not specially on annealing conditions..

To investigate the physics underlying the origin of the reduction and subsequent inversion of *S*, we measured the diamagnetic shifts of the X and XX states for the same quantum dots of figure 2 to probe the extent of the exciton wavefunction in the plane of the quantum dots. Magnetic fields were generated using a 5T superconducting magnet. The samples were studied in the Faraday geometry (field perpendicular to the plane). In this geometry the Zeeman interaction increases the splitting of the X and XX doublets, and diamagnetic confinement increases the average X and XX energies with increasing field. Exciton (biexciton) diamagnetic shifts were measured for a large number of dots in all the samples by measuring the change in the average energy of the exciton (biexciton) doublet with increasing field. The diamagnetic shifts were found to be proportional to the square of the applied field as expected[23]; the coefficients of proportionality for each dot are plotted in figure 3 as a function of the corresponding dot emission energy. On average the biexciton diamagnetic shift was found to be around $3 \pm 1$ µeV/T$^{-2}$ smaller than that of the exciton, demonstrating the importance of the balance between magnetic confinement and coulomb repulsion in the biexciton state.

We observe increasing diamagnetic shift with increasing emission energy for both X and XX emission, in contrast to what one expects for strongly confining quantum dots of decreasing dimensions. Instead, these results demonstrate that the size of the exciton envelope is increasing with emission energy, which we attribute to weakly confined exciton states, which increasingly tend to extend into the barrier material as the dimensions of the quantum dot reduce. The in plane expansion of the exciton, a consequence of the reduction of the confinement potential, causes a reduction in the long range exchange interaction, and hence a reduction in $S$[24]. We note that the blue shift of the wetting layer on annealed samples is only up to ~10meV, and has negligible effect on the confinement energy.

The reduction in confinement and subsequent expansion of the exciton envelope[24] provides important insight into the results presented here. Crucially however, a simple interpretation of this mechanism would be that the splitting decreases as the confinement reduces, with $S$ tending to zero as confinement is lost. Such a situation, anticipated by studies of the splitting of ensembles of quantum dots[18], would have limited usefulness due to the conflicting requirements of weak confinement, and strong radiative recombination. However, the inversion of $S$ presented above, impossible to detect in previous experiments that probe the magnitude of the splitting only, has strong potential to allow the selection of reasonably confined quantum dots with no exciton level splitting, for which radiative processes still dominate. We believe the most likely cause of the inversion of $S$ lies with subtle interplay between piezoelectric and shape induced exciton asymmetry. It is well known that InAs quantum dots tend to elongate along [1-10][12], thus elongating exciton states in that direction. Piezoelectric potentials however, caused by shear strain from the lattice mismatch between InAs and GaAs, tend to elongate the hole in the [1-10] direction, and the electron in the [110] direction[25]. Thus for strongly confined states, shear strain tends to also elongate the exciton in the [1-10] direction, since the extent of the exciton wavefunction is dominated by the small hole. However, as confinement is reduced, shape asymmetry becomes less dominant as the electron and hole states extend into the barrier, and the relative size and elongation of the electron and hole states could potentially lead to an exciton state elongated along [110], the preferred axis of expansion for the electron.

Although the above results indicate that annealing is not necessary to tune the splitting, since it can be varied equally well by growth alone, during the course of these measurements we have observed a number of annealing specific effects, most of which present advantages. Longer anneal times and higher anneal temperatures do make the selection of higher energy dots easier. Vertical diffusion makes the dots shallower, increases their emission energies and therefore reduces the number of low energy dots found; there were no dots found to be optically active below 1.38eV in the sample annealed for 10 minutes at 750°C in the larger number we studied for example. Annealing thus presents an efficient way to predictably vary the emission energy and dot density, with more control than by growth alone.

The most striking feature of annealed quantum dots is revealed in Fig. 4, which plots the biexciton binding energy ($E_X - E_{XX}$) for a number of annealed and un-annealed quantum dots. We observe a clear transition from negative biexciton binding energies for the un-annealed sample (where the biexciton emission is blue-shifted relative to the exciton emission) to only positive binding energies for the sample annealed at the highest temperature (where biexciton emission is red-shifted). The measured increase in binding energy with annealing has been associated with increased symmetry in the vertical potential profile of the dot after annealing[18], decreasing the electron-hole separation, and increasing Coulomb attraction. Such an increase in the electron-hole overlap also implies increased oscillator strength[26], which could be very important to allow radiative recombination to compete favourably against non radiative carrier escape processes in these shallow dots. We note that we do not observe any clear relationship between the binding energy of the biexciton and the exciton recombination energy observed elsewhere[27], due to the large scatter on the measured binding energies, and the smaller range of quantum dot emission energies. The range of binding energies measured however, are similar.

For the emission of entangled photon pairs to be possible, then |S| must be within the homogeneous linewidth of the emission, determined to be ~1µeV from the radiative lifetime of these quantum dots. Amongst the dots studied here, four were measured to have the required |S|<0.5±0.5µeV, which is extremely encouraging. In any case, the fact that S can be reduced close to zero, and inverted strongly suggests that exciton degeneracy can be achieved by growth or annealing, with annealing increasing the number of suitable dots. We note that the conditions required for degeneracy can be further relaxed due to Purcell enhancement of the homogenous linewidth in optical cavities, where Purcell factors of ~5 can be achieved[5,28] corresponding to homogeneous linewidths of ~5µeV.

In conclusion, we have presented measurements of the fine structure splitting of the exciton level for a large number of single InAs quantum dots, revealing a switching of the polarisation of the doublet with increasing recombination energy. Measurements of the exciton diamagnetic shift in the Faraday geometry as a function of the average recombination energy suggest that the exciton's wavefunction loses confinement in the higher energy dots, allowing strain induced elongation of electrons along [110], to be the dominant contribution to the asymmetry of the exciton envelope. Annealing has been shown to select higher energy dots for individual study, though not to directly change the splitting of the exciton level as has previous been suggested. Finally, the wavelength dependent transition of the exciton splitting through zero is of crucial importance, and represents the first step in allowing single quantum dots to be used as a source of polarisation-entangled photon pairs.

We would like to acknowledge continued support from Prof. M. Pepper. This work was partially funded by the EPSRC IRC for Quantum Information Processing, EPSRC grant GR/S76076, and the EU Framework Package 6 Network of Excellence SANDiE.

**Figure Captions**

**Fig. 1**: Photoluminescence spectra from two single quantum dots; (a) from an un-annealed sample and (b) from a sample annealed for five minutes at 750°C. Solid and dotted lines represent horizontally (H) and vertically (V) polarised detection for (a) and (b) respectively. Two groups of lines are observed corresponding to emission from exciton (X) and biexciton (XX) states as labelled, the inset to the figure shows a simple energy level diagram illustrating optical decay from the XX state, the exciton level splitting S is labelled.

**Fig. 2**: Exciton level splitting S (defined in the text, is positive when the horizontally polarised component is to higher energy than the vertically) as a function of the exciton recombination energy for a large number of single quantum dots measured on annealed and un-annealed samples. The dotted line shows a linear fit to the data and has a gradient of $-(1.3 \pm 0.1) \times 10^{-3}$. The histogram inset shows the measured error on the splitting.

**Fig 3**: Diamagnetic shifts of exciton (X, right axis) and biexciton (XX, left axis) states in single quantum dots as a function of the exciton recombination energy. Data for annealed and un-annealed samples is shown.

**Fig 4**: Biexciton binding energy as a function of the exciton recombination energy for a large number of single quantum dots measured on annealed and un-annealed samples.

**Fig1:**

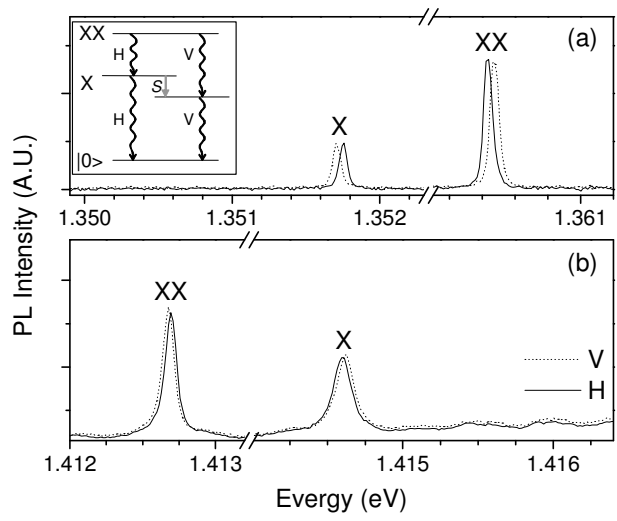

**Fig2:**

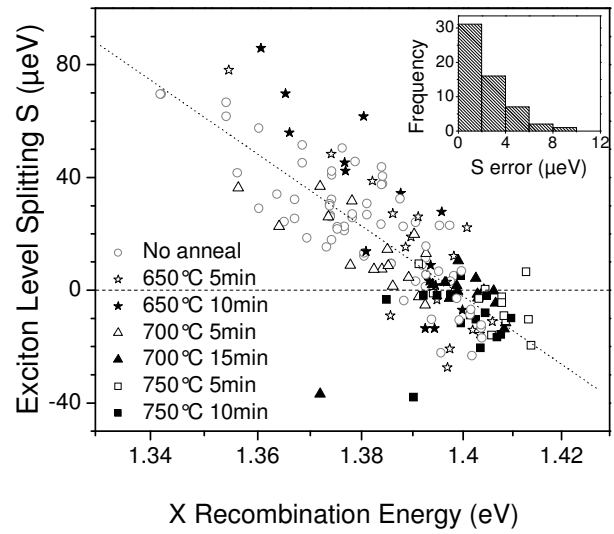

**Fig3:**

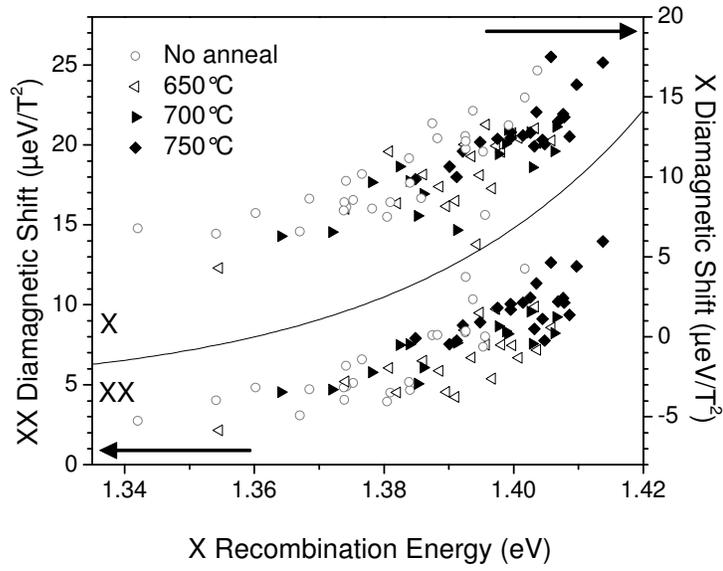

**Fig4:**

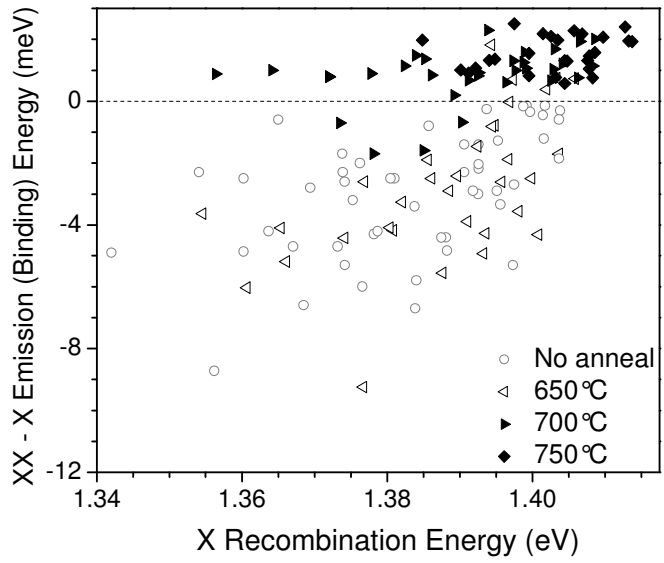

**References**

[1] P. Michler, A. Kiraz, C. Becher, W. V. Schoenfeld, P. M. Petroff, Lidong Zhang, E. Hu, and A. Imamoglu, Science **290**, 2282 (2000).

[2] C. Santori, M. Pelton, G. Solomon, Y. Dale, and Y. Yamamoto, Phys. Rev. Lett. **86**, 1502 (2001).

[3] R. M. Thompson, R. M. Stevenson, A. J. Shields, I. Farrer, C. J. Lobo, D. A. Ritchie, M. L. Leadbeater, and M. Pepper, Phys. Rev. B **64**, 201302(R) (2001).

[4] Z. Yuan, B. E. Kardynal, R. M. Stevenson, A. J. Shields, C. J. Lobo, K. Cooper, N. S. Beattie, D. A. Ritchie, and M. Pepper, Science **295**, 102 (2002).

[5] M. Pelton, C. Santori, J. Vuckovic, B. Zhang, G. S. Solomon, J. Plant, and Y. Yamamoto, Phys. Rev. Lett. **89**, 233602 (2002).

[6] G. S. Solomon, M. Pelton, and Y. Yamamoto, Phys. Rev. Lett. **86**, 3903 (2001).

[7] C. Santori, D. Fattal, J. Vučković, G.S. Solomon, and Y. Yamamoto, Nature **419**, 594 (2002).

[8] A. J. Bennett, D. C. Unitt, P. Atkinson, D. A. Ritchie, and A. J. Shields, Opt. Express **13**, 50-55 (2005).

[9] D. Fattal, K. Inoue, J. Vuckovic, C. Santori, G. S. Solomon, and Y. Yamamoto., Phys. Rev. Lett. **92**, 037903 (2004).

[10] E. Moreau, I. Robert, L. Manin, V. Thierry-Mieg, J. M. Gérard, and I. Abram, Phys. Rev. Lett. **87**, 183601 (2001).

[11] O. Benson, C. Santori, M. Pelton, and Y. Yamamoto, Phys. Rev. Lett. **84**, 2513 (2000).

[12] R. M. Stevenson, R. M. Thompson, A. J. Shields, I. Farrer, B. E. Kardynal, D. A. Ritchie, and M. Pepper, Phys. Rev. B **66**, 081302(R) (2002).

[13] C. Santori, D. Fattal, M. Pelton, G.S. Solomon, and Y. Yamamoto Phys. Rev. B **66**, 045308 (2002).

[14] S. M. Ulrich, S. Strauf, P. Michler, G. Bacher, and A. Forchel, Appl. Phys. Lett. **83**, 1848 (2003).

[15] V. D. Kulakovskii, G. Bacher, R. Weigand, T. Kümmell, A. Forchel, E. Borovitskaya, K. Leonardi, and D. Hommel, Phys. Rev. Lett. **82**, 1780 (1999).

[16] J. Persson, M. Holm, C. Pryor, D. Hessman, and W. Seifert, Phys. Rev. B **67**, 035320 (2003)

[17] D. Gammon, E.S. Snow, B. V. Shanabrook, D. S. Katzer, and D. Park, Phys. Rev. Lett. **76**, 3005 (1996).

[18] W. Langbein, et al., Phys. Rev. B **69**, 161301(R) (2004).

[19] Tartakovskii et al, to be published in PRB.

[20] M. Geiger, A. Bauknecht, F. Adler, H. Schweizer, and F. Scholz, J. Cryst. Growth **170,** 558 (1997).

[21] V. D. Kulakovskii, et al., Phys. Rev. Lett. **82**, 1780 (1999).

[22] M. Bayer, and A. Forchel, Phys. Rev. B **65**, 041308(R) (2002).

[23] S. N. Walck, and T. L. Reinecke, Phys. Rev. B **57**, 9088 (1998).

[24] S. Fafard, and C.Nì. Allen, Appl. Phys. Lett. **75**, 2374 (1999)

[25] O. Stier, M. Grundmann, and D. Bimberg, Phys. Rev. B **59**, 5688 (1999).

[26] W. Langbein, et al., Phys. Rev. B **70**, 033301 (2004).

[27] S. Rodt, et al., Phys. Rev. B **68**, 035331 (2003).

[28] A. J. Bennett, et al, Optics Express 13, 50 (2005)